\documentclass[acmsmall]{acmart}

\AtBeginDocument{%
  \providecommand\BibTeX{{%
    \normalfont B\kern-0.5em{\scshape i\kern-0.25em b}\kern-0.8em\TeX}}}

\usepackage{times}

\usepackage{soul}
\usepackage{url}
\usepackage{graphicx}
\usepackage{amsmath}
\usepackage{booktabs}
\usepackage{array}
\usepackage{amsthm}
\usepackage{multirow}
\usepackage{graphicx}
\usepackage{subfloat}
\DeclareMathAlphabet{\mathpzc}{OT1}{pzc}{m}{it}

\begin{document}
\title{Deep Learning on Knowledge Graph for Recommender System: A Survey}

\author{Yang Gao}
\authornote{Both authors contribute equally to this research.}
\email{yxg122530@utdallas.edu}
\author{Yi-Fan Li}
\authornotemark[1]
\email{yli@utdallas.edu}
\affiliation{%
  \institution{University of Texas at Dallas}
}
\author{Yu Lin}
\affiliation{%
  \institution{University of Texas at Dallas}
}
\email{yxl163430@utdallas.edu}

\author{Hang Gao}
\affiliation{%
  \institution{University of Maryland Baltimore County}
}
\email{hanggao1@umbc.edu}

\author{Latifur Khan}
\affiliation{%
  \institution{University of Texas at Dallas}
}
\email{lkhan@utdallas.edu}







\begin{abstract}
Recent advances in research have demonstrated the effectiveness of knowledge graphs (KG) in providing valuable external knowledge to improve recommendation systems (RS). 
A knowledge graph is capable of encoding high-order relations which connect two objects with one or multiple related attributes.
With the help of the emerging Graph Neural Networks (GNN), it is possible to extract both object characteristics and relations from KG, which is an essential factor for successful recommendations.
In this paper, we provide a comprehensive survey of the GNN-based knowledge-aware deep recommender systems. 
Specifically, we discuss the state-of-the-art frameworks with a focus on their core component, i.e., the graph embedding module, and how they address practical recommendation issues such as scalability, cold-start and so on.
We further summarize the commonly-used benchmark datasets, evaluation metrics as well as the open-source codes. Finally, we conclude the survey and propose potential research directions in this rapidly growing field.

\end{abstract}




\maketitle

\section{Introduction}

In modern life, recommender systems (RS) are critical for users to make proper choices from the huge amount of products or services.
For example, people are lured in AI-driven recommendation service for more than $75\%$ of the time they spent on watching YouTube videos~\cite{solsman2018}.
These systems attempt to learn the users' interests and keep them away from over-choice, which significantly boosts their decision making process~\cite{jannach2010recommender}.
They also help to promote services and sales for business such as e-commerce.

\begin{figure}[t]
\centering
\includegraphics[width=0.7\columnwidth]{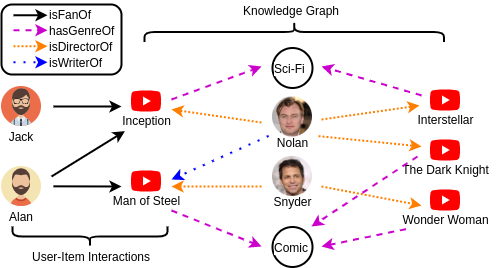}
\caption{Illustration of user-item interactions and knowledge graph on YouTube, which contains users, movies, directors and genres as objects; liking, categorizing, directing and writing as object relations}
\label{fig:teasing_figure}
\end{figure}

There are two main architectures for traditional recommender systems: content-based system and collaborative-filtering based system. Content-based systems accept data that can be represented as vectors in an euclidean space~\cite{DBLP:journals/corr/abs-1901-00596} and measure their similarities based on these vectors. Collaborative-filtering based systems usually assume that each user-item interaction is an independent instance with side information encoded~\cite{DBLP:conf/kdd/Wang00LC19}.
However, in modern society, there is an increasing number of applications where data are generated from non-Euclidean domains and are represented in the form of Knowledge Graphs (KG). This data structure breaks the independent interaction assumption by linking items with their attributes.
For example, in YouTube RS (show in Fig.~\ref{fig:teasing_figure}), the connections between users and movies might be actions such as views, likes and comments. On the other hand, movies may share the same genre, director, and etc. 
With these information, a knowledge-aware RS is hence able to capture not only the user-item interactions but also the rich item-item/user-user relations to make more accurate recommendations. It is worth noticing that nodes in the KG  (e.g., person or movies) may have distinct neighborhood size (i.e., number of neighboring nodes), and the relationships between them could vary as well, which makes recommendation with KG even more challenging.

Recently, there is an increasing interest in extending deep learning approaches for graph data. Motivated by CNNs, RNNs, and autoencoders from deep learning, new neural network architectures have been rapidly developed over the past few years to handle the complexity of graph data. These types of networks are referred as Graph Neural Networks (GNNs)~\cite{DBLP:journals/corr/abs-1901-00596} and are critical for recent success of knowledge-aware deep recommender systems.

There are a number of existing reviews on the topic of graph neural networks and knowledge graph analysis. Wu et al.~\cite{DBLP:journals/corr/abs-1901-00596} give an overview of  graph neural networks in data mining and machine learning fields. Although it is the first comprehensive review of GNNs, this survey mainly focuses on the network architectures and only briefly mentions the possibility of applying GNNs for recommendation. On the other hand, Shi et al.~\cite{DBLP:journals/tkde/ShiLZSY17} summarize the recommender systems utilizing the traditional knowledge graph analysis methods. But it misses the most recent development of GNN-based algorithms. To our best knowledge, our survey is the first review of state-of-the-art deep recommender systems that utilize GNNs to extract information from knowledge graphs for improving recommendation.

\textbf{Our Contributions:} Our paper makes notable contributions summarized as follows:
\begin{itemize}
\item \textbf{New Taxonomy:} The core component of knowledge-aware deep recommender systems is the graph embedding module, which is usually a GNN in the state-of-the-art frameworks. Each layer of a GNN consists of two basic components: Aggregator and Updater. We categorize the aggregators into three groups: relation-unaware aggregator, relation-aware subgraph aggregator, and relation-aware attentive aggregator. We also divide the updaters into three categories, i.e., context-only updater, single-interaction updater and multi-interaction updater.
\item \textbf{Comprehensive Overview:} We provide the most comprehensive overview of state-of-the-art \textit{GNN-based Knowledge Aware Deep Recommender} (\textbf{GNN-KADR}) systems. We provide detailed descriptions on representative models, make the necessary comparison, and discuss their solution to practical recommendation issues such as cold start, scalability and so on.
\item \textbf{Resource Collection:} We summarize resources regarding GNN-KADR systems, including benchmark datasets, evaluation metrics and open-source codes.
\item \textbf{Future Directions:} We analyze the limitations of existing works and suggest a few future research directions such as dynamicity, interpretability, fairness and so on.
\end{itemize}

\textbf{Organization of Our Survey: }
The rest of the survey is organized as follows. Section~\ref{sec:preliminary} defines the concepts related to GNN-based knowledge aware recommendations and lists the notations used in this paper. Section~\ref{sec:categorization} provides an overview of state-of-the-art GNN-KADR systems. Section~\ref{sec:issues} discusses their solutions to practical recommendation issues. Section~\ref{sec:evaluations} outlines the widely used benchmark datasets, evaluation metrics, and open-source codes. Section~\ref{sec:future} discusses the current challenges and suggests future research directions.  Section~\ref{sec:conclusions} summarizes the paper.

\label{sec:introduction}

\section{Preliminary and Notation}
\begin{figure*}[t]
\centering
\includegraphics[width=0.9\textwidth]{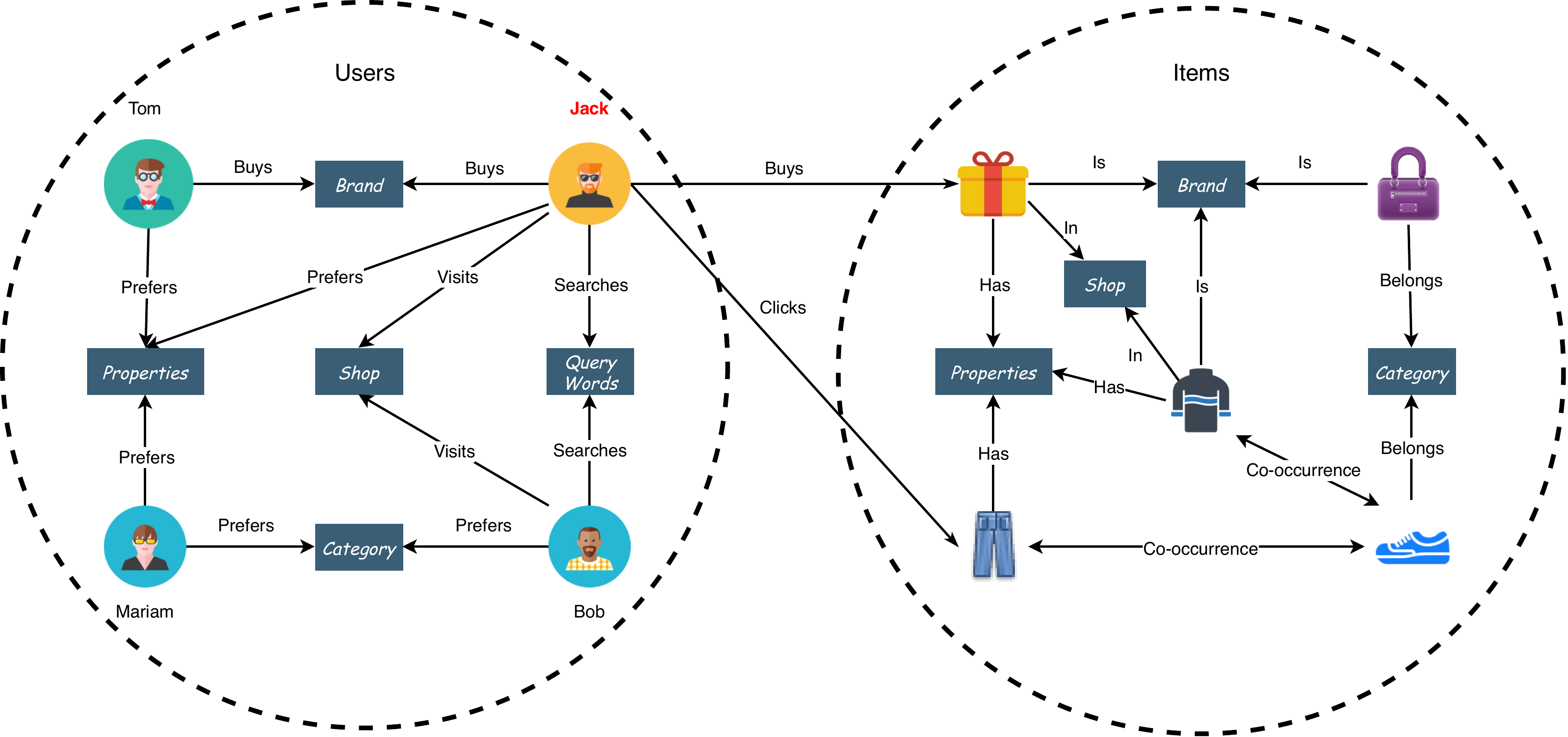}
\caption{Example of a knowledge graph on e-commerce websites.}
\label{fig:kg_example}
\end{figure*}

First, let us consider a typical scenario on a e-commerce website: Jack is a customer who prefers to wear blue jeans and has queried about ``jeans'' one day. From the list of returned items, Jack clicked some attractive items for detailed information. During this week, he also visited some online shops for checking out T-shirts. Finally, on Sunday, Jack purchased a blue jean from his favorite brand as a birthday gift and added another T-shirt in the same shop to his shopping cart. Based on Jack's behavior, the platform has collected rich information (submitted query words, clicked items, visited shops, preferred properties and brands) for recommending potential interesting items to him in the future. This kind of recommendation scenarios could also be observed on other websites.
In general, multiple kinds of objects and historical user behaviors form a knowledge graph.
Figure~\ref{fig:kg_example} shows a toy example on e-commerce websites. 


\begin{definition}
\textit{\textbf{Knowledge Graph} (KG)~\cite{DBLP:conf/recsys/Sun00BHX18}} is defined as a \textit{directed} graph $\mathcal{G}=(\mathcal{V}, \mathcal{E})$, where $\mathcal{V}$ is the set of nodes and $\mathcal{E} \subseteq \mathcal{V} \times \mathcal{V}$ is the set of edges between nodes in $\mathcal{V}$. $\mathcal{G}$ is associated with a node type mapping function $\phi$: $\mathcal{V} \rightarrow \mathcal{A}$ and an edge type mapping function $\psi$: $\mathcal{E} \rightarrow \mathcal{R}$, where $|\mathcal{A}|>1$ and/or $|\mathcal{R}|>1$. Each node $v\in \mathcal{V}$ belongs to one particular node type in the node type set $\mathcal{A}$: $\phi(v)\in \mathcal{A}$, and each edge $e \in \mathcal{E}$ belongs to a particular relation type in relation type set $\mathcal{R}$: $\psi(e) \in \mathcal{R}$. 
Mining knowledge graphs is usually based on a basic entity-relation-entity triplet $(u, \mathpzc{e}, v)$, where $u\in \mathcal{V}$, $\mathpzc{e}\in \mathcal{E}$, $v\in \mathcal{V}$ denote the head, relation and tail of this triplet. In this paper, we refer these entity-relation-entity triplets as \textit{knowledge triplets} for simplicity. Here the types of nodes $u$ and $v$ could be either same or different depending on the context.
\end{definition}

\begin{definition}
\textit{\textbf{Neighborhood} $N(v)$.} Given a knowledge graph $\mathcal{G}=(\mathcal{V}, \mathcal{E})$, for a node $v$, its neighborhood $N(v)$ is defined as the set of nodes that directly connect to $v$, i.e., $\{w|(w,\mathpzc{e},u)\textit{ or } (u,\mathpzc{e},w), \mathpzc{e}\in \mathcal{E}\}$.
\end{definition}

\begin{definition}
\textit{\textbf{$r$-Neighborhood} $N_r(v)$.} Given a knowledge graph $\mathcal{G}=(\mathcal{V}, \mathcal{E})$, for a node $v$, its $r$-neighborhood $N_r(v)$ is defined as the set of nodes that connect to $v$ with edges of type $r$, i.e., $\{w|(w,\mathpzc{e},u)\textit{ or } (u,\mathpzc{e},w), \textit{ where } \mathpzc{e}\in \mathcal{E} \textit{ and } \psi(\mathpzc{e})=r\}$.
\end{definition}

\noindent
\textbf{User-Item Recommendation} In general, $\mathcal{V}$ can be further written as $\mathcal{V}=\mathcal{V}_1\cup \mathcal{V}_2\cup \cdots \cup \mathcal{V}_i \cup \cdots \cup \mathcal{V}_n$, where $\mathcal{V}_i$ denotes the set of nodes with type $i$ and $n=|\mathcal{A}|$. In recommendation we denote $\mathcal{V}_1$ as the set of user nodes and $\mathcal{V}_2$ as the set of item nodes, with $\mathcal{V}_3, \ldots, \mathcal{V}_n$ representing the other objects' nodes (brand, properties, etc). We also denote $\mathcal{E} = \mathcal{E}_{label}\cup \mathcal{E}_{unlabel}$, where $\mathcal{E}_{label} \subseteq \mathcal{V}_1\times \mathcal{V}_2$ represents the set of edges between user and item nodes, and $\mathcal{E}_{unlabel}=\mathcal{E} \setminus \mathcal{E}_{label}$ represents other edges. Since a typical recommendation setting in real world is to predict a user's preferred items based on his historical behaviors, we use $\mathcal{G}=(\mathcal{V}, \mathcal{E})$ to denote the graph constructed from historical data, and $\mathcal{G}^P=(\mathcal{V}^P, \mathcal{E}^P)$ to denote the graph of real future. The user-item recommendation problem can hence be formulated as a link prediction problem on graph:

\textbf{Input:} A KG $\mathcal{G}=(\mathcal{V}, \mathcal{E})$ constructed based on historical data.

\textbf{Output:} A predicted edge set $\widehat{\mathcal{E}}^p_{label}$, which is the prediction of the real edge set $\mathcal{E}^P_{label}$ on $\mathcal{G}^P$.

\bigskip

Throughout this paper, the bold uppercase characters are used to denote matrices and bold lowercase characters are used to denote vectors. Unless particularly specified, the notations used in this paper are illustrated in Table~\ref{tab:notations}. 

\begin{table}[t]
\centering
\caption{Commonly used notations.}
\label{tab:notations}
\resizebox{0.9\columnwidth}{!}{%
\begin{tabular}{ll}
\toprule[2pt]
\multicolumn{1}{l}{\textbf{Notations}} & \multicolumn{1}{l}{\textbf{Descriptions}} \\
\specialrule{0.5pt}{2pt}{2pt} 
$|\cdot|$ & The size of a set. \\
\specialrule{0.5pt}{2pt}{2pt}
$\odot$ & Element-wise product. \\
\specialrule{0.5pt}{2pt}{2pt}
$||$ & Concatenation\\
\specialrule{0.5pt}{2pt}{2pt}
$\mathcal{G}$ & A knowledge graph. \\
\specialrule{0.5pt}{2pt}{2pt}
$\mathcal{V}$ & The set of nodes in a knowledge graph. \\
\specialrule{0.5pt}{2pt}{2pt}
$\mathcal{E}$ & The set of edges between nodes in $\mathcal{V}$.\\
\specialrule{0.5pt}{2pt}{2pt}
$\mathcal{A}$ & The node type set.\\
\specialrule{0.5pt}{2pt}{2pt}
$\mathcal{R}$ & The edge relation type set.\\
\specialrule{0.5pt}{2pt}{2pt}
$v$ & A node $v \in \mathcal{V}$. \\
\specialrule{0.5pt}{2pt}{2pt}
$e_{i,j}$ & An edge $e_{i,j}\in \mathcal{E}$. \\
\specialrule{0.5pt}{2pt}{2pt}
$(u, \mathpzc{e}, v)$ & A knowledge triplet where $u\in \mathcal{V}$, $\mathpzc{e}=e_{u,v}\in \mathcal{E}$, $v\in \mathcal{V}$ denote the head, relation and tail of this triplet. \\
\specialrule{0.5pt}{2pt}{2pt}
$\mathbf{z}_v\in \mathbf{R}^d$ & The feature vector of node $v$\\
\specialrule{0.5pt}{2pt}{2pt} 
$\mathbf{z}_{\mathpzc{e}}\in \mathbf{R}^c$ & The feature vector of edge $\mathpzc{e}=e_{u,v}$ or $e_{v,u}$\\
\specialrule{0.5pt}{2pt}{2pt} 
$\mathbf{z}_{\mathpzc{e}_{i,j}}\in \mathbf{R}^c$ & The feature vector of edge $e_{i,j}$\\
\specialrule{0.5pt}{2pt}{2pt} 
$\mathbf{n}_u\in \mathbf{R}^k$ & The context representation of node $u$\\
\specialrule{0.5pt}{2pt}{2pt} 
$\mathbf{n}_u^r\in \mathbf{R}^k$ & Type-$r$ context representation of node $u$\\
\specialrule{0.5pt}{2pt}{2pt} 
$\mathbf{X}\in \mathbf{R}^{N\times d}$ & The feature matrix of nodes in a knowledge graph.\\
\specialrule{0.5pt}{2pt}{2pt}
$\mathbf{A} \in \mathbf{R}^{N\times N}$ & The adjacency matrix of a knowledge graph \\
\specialrule{0.5pt}{2pt}{2pt}
$\mathbf{N} \in \mathbf{R}^{N\times k}$ & The context representations of all nodes in a knowledge graph \\
\specialrule{0.5pt}{2pt}{2pt}
$N(v)$ & The neighborhood of a node $v$ \\
\specialrule{0.5pt}{2pt}{2pt} 
$N_r(v)$ & $r$-Neighborhood of a node $v$. \\
\specialrule{0.5pt}{2pt}{2pt}
$n$ & The number of nodes, $n=|\mathcal{V}|$. \\
\specialrule{0.5pt}{2pt}{2pt}
$m$ & The number of edges, $m=|\mathcal{E}|$. \\
\specialrule{0.5pt}{2pt}{2pt}
$d$ & The dimension of a node feature vector. \\
\specialrule{0.5pt}{2pt}{2pt}
$c$ & The dimension of an edge feature vector \\
\specialrule{0.5pt}{2pt}{2pt} 
$k$ & The dimension of $v$'s context representation. \\
\specialrule{0.5pt}{2pt}{2pt}
$\sigma(\cdot)$ & The sigmoid activation function \\
\specialrule{0.5pt}{2pt}{2pt} 
$\beta(\cdot)$ & The LeakyReLU activation function\\
\specialrule{0.5pt}{2pt}{2pt} 
$\gamma(\cdot)$ & An activation function, e.g., sigmoid, ReLU, etc.\\
\specialrule{0.5pt}{2pt}{2pt} 
$\mathbf{W}$, $\mathbf{\Theta}$, $\mathbf{b}$ & Learnable model parameters. \\

\bottomrule[2pt]
\end{tabular}%
}
\end{table}

\label{sec:preliminary}

\section{Categorization and Frameworks}
\label{sec:categorization}
\begin{figure}[t]
\centering
\includegraphics[width=0.9\columnwidth]{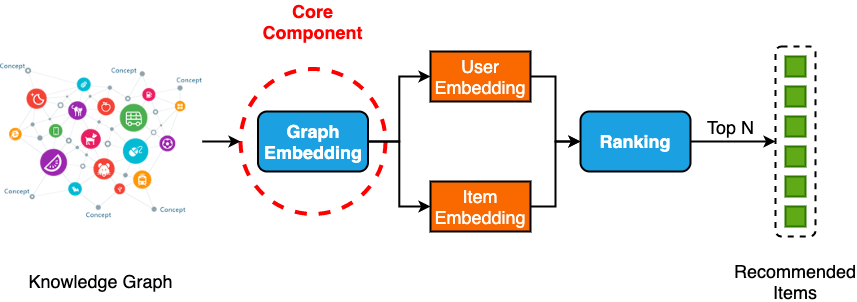}
\caption{The general workflow of a GNN-based Knowledge Aware Deep Recommender (GNN-KADR) system.}
\label{fig:kg_workflow}
\end{figure}

The general workflow of a GNN-based Knowledge Aware Deep Recommender (GNN-KADR) system is shown in Figure~\ref{fig:kg_workflow}. 
The graph embedding module of GNN-KADR first learns to produce an embedding for every graph node (including the user and item nodes), which encodes the information distilled from the input knowledge graph.
Next, for a given user, the ranking module computes a match score between this user and each of the candidate items according to their corresponding embeddings. Those items with top-N match scores (or match scores above a user-specified threshold) are linked (recommended) to this user.

In this process, the core component of the system is the graph embedding module, which is usually a Graph Neural Network (GNN) in the state-of-the-art frameworks. Each layer of a GNN consists of two basic components: Aggregator and Updater. For a node $v$, the Aggregator aggregates the feature information from the neighbors of $v$ to produce a context representation. The Updater then utilizes this context representation as well as other input information to obtain a new embedding for node $v$. Stacking $K$ different GNN layers or reusing the same GNN layer $K$ times expands GNN's receptive field to $K$-hop graph neighborhoods.

\begin{figure}[]
\centering
\includegraphics[width=0.8\columnwidth]{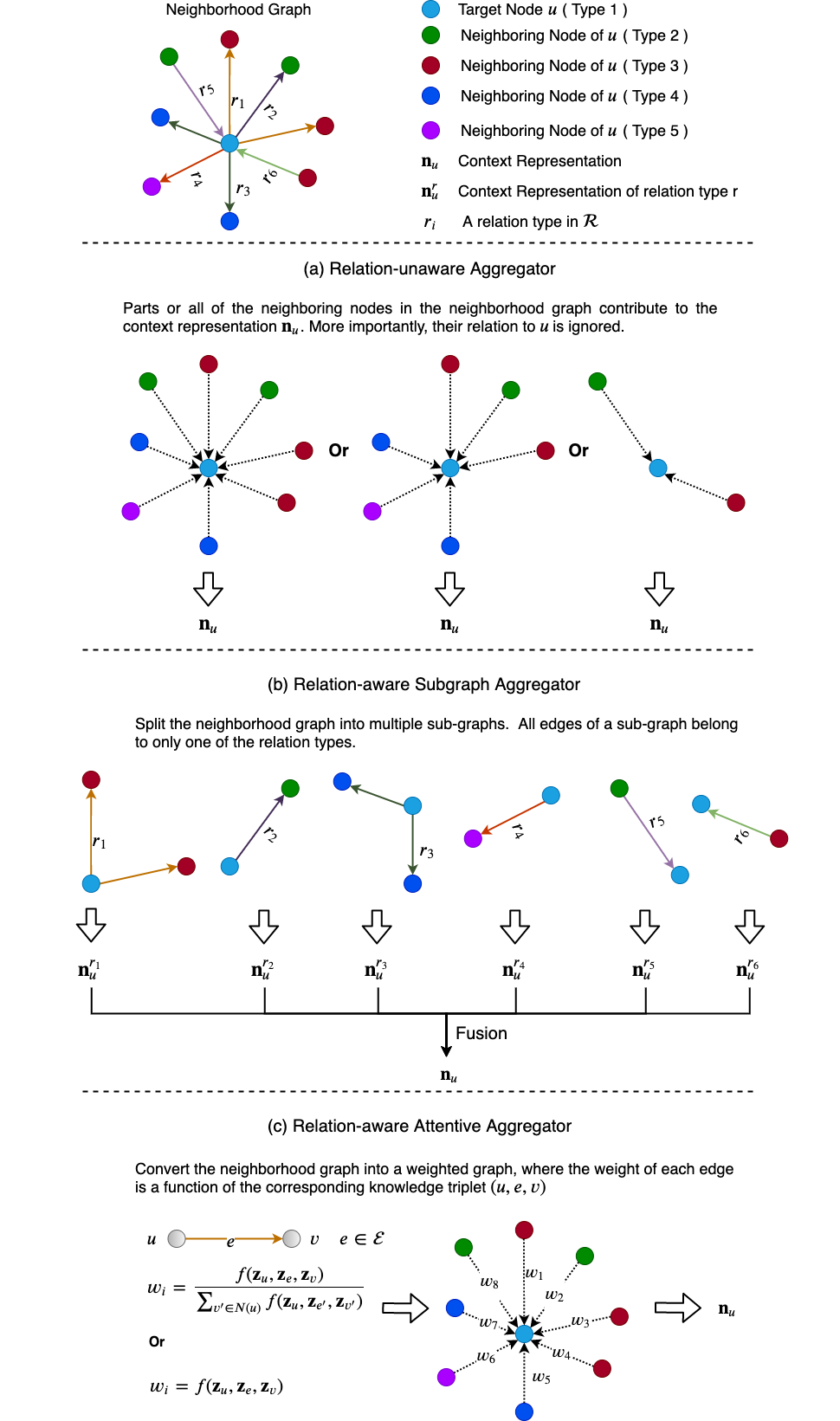}
\caption{Description of relation-unaware aggregator, relation-aware subgraph aggregator, and relation-aware attentive aggregator.}
\label{fig:aggregator}
\end{figure}

\begin{figure}[]
\centering
\includegraphics[width=\columnwidth]{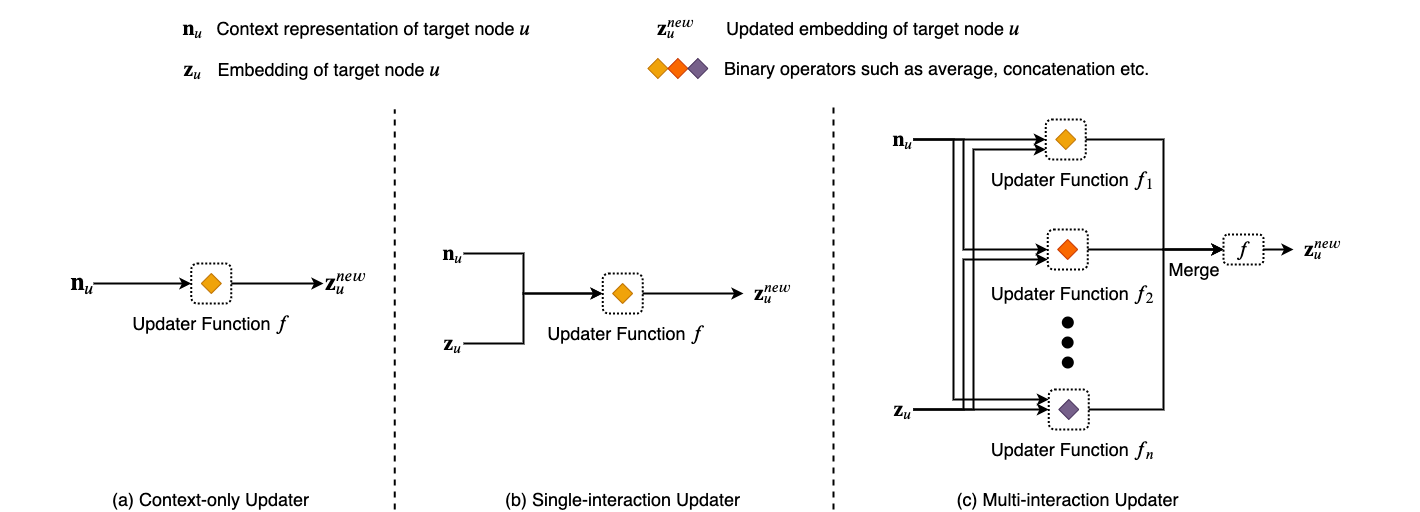}
\caption{Description of context-only updater, single-interaction updater, and multi-interaction updater.}
\label{fig:updater}
\end{figure}

We categorize the aggregators of GNNs into relation-unaware aggregator, relation-aware subgraph aggregator and relation-aware attentive aggregator. On the other hand, we divide the updaters into three categories, i.e., context-only updater, single-interaction updater and multi-interaction updater.
Various categories of aggregators and updaters are described in Figure~\ref{fig:aggregator} and Figure~\ref{fig:updater} respectively.
In the following, we discuss them in details.

\begin{table}[t]
\centering
\caption{Taxonomy and representative publications of GNN-based Knowledge Aware Deep Recommender (GNN-KADR) systems.}
\label{tab:taxonomy}
\resizebox{\textwidth}{!}{%
\begin{tabular}{llll}
\toprule[2pt]
\textbf{GNN} & \textbf{Category} & \textbf{} & \multicolumn{1}{c}{\textbf{Publications}} \\ \hline
\multirow{3}{*}{Aggregator} & Relation-unaware Aggregator &  & ~\cite{fan2019metapath},~\cite{ying2018graph} \\ \cline{2-4} 
 & \multirow{2}{*}{Relation-aware Aggregator} & Subgraph Aggregator & ~\cite{xu2019relation},~\cite{zhang2019star}, ~\cite{zhao2019intentgc} \\ \cline{3-4} 
 &  & Attentive Aggregator & ~\cite{fan2019graph},~\cite{li2019long},~\cite{DBLP:conf/wsdm/Song0WCZT19},~\cite{wang2019knowledge},~\cite{DBLP:conf/www/WangZXLG19},~\cite{wang2019kgat},~\cite{wu2019dual} \\ \hline
\multirow{3}{*}{Updater} & Context-only Updater &  & ~\cite{fan2019metapath},~\cite{fan2019graph},~\cite{li2019long},~\cite{DBLP:conf/wsdm/Song0WCZT19},~\cite{DBLP:conf/www/WangZXLG19},~\cite{wu2019dual},~\cite{zhang2019star} \\ \cline{2-4} 
 & Single-interaction Updater &  & ~\cite{wang2019knowledge},~\cite{DBLP:conf/www/WangZXLG19},~\cite{xu2019relation},~\cite{ying2018graph},~\cite{zhao2019intentgc} \\ \cline{2-4} 
 & Multi-interaction Updater &  & ~\cite{wang2019kgat}\\
 \bottomrule[2pt]
\end{tabular}%
}
\end{table}

\begin{table}[]
\centering
\caption{Summary of studied GNN-KADR systems.}
\label{tab:framework_summary}
\resizebox{\textwidth}{!}{%
\begin{tabular}{llll}
\toprule[2pt]
\multicolumn{1}{l}{\textbf{Method}} & \multicolumn{1}{l}{\textbf{Aggregator}} & \multicolumn{1}{l}{\textbf{Updater}} & \multicolumn{1}{l}{\textbf{Issues Solved}} \\
\specialrule{0.5pt}{2pt}{2pt} 
MEIRec~\cite{fan2019metapath} & $\mathbf{n}_u=Avg/LSTM/CNN(\{\mathbf{z}_v|v\in N(u)\})$ & $\mathbf{z}_u^{new}=\mathbf{n}_u$ & Scalability; Cold-Start \\
\specialrule{0.5pt}{2pt}{2pt} 

\multirow{4}{*}{GraphRec~\cite{fan2019graph}} & $\mathbf{x}_{u,v}=MLP(\mathbf{z}_v||\mathbf{z}_{\mathpzc{e}})$ & \multirow{4}{*}{$\mathbf{z}_u^{new}=\gamma(\mathbf{W}\cdot \mathbf{n}_u+\mathbf{b})$} & \multirow{4}{*}{---} \\
 & $\alpha_{u,v}^*=\mathbf{W}_2^T\cdot \gamma(\mathbf{W}_1\cdot (\mathbf{x}_{u,v}||\mathbf{z}_u)+\mathbf{b}_1)+\mathbf{b}_2$ &  &  \\
 & $\alpha_{u,v} = \frac{\exp(\alpha_{u,v}^*)}{\sum_{v'\in N(u)}\exp(\alpha_{u,v'}^*)}$ &  &  \\
 & $\mathbf{n}_u=\sum_{v\in N(u)} \alpha_{u,v} \mathbf{x}_{u,v}$ &  & \\
 \specialrule{0.5pt}{2pt}{2pt}

\multirow{3}{*}{V2HT~\cite{li2019long}} & $\mathbf{A}_{i,j}=f_{weight}(\mathbf{X}[i,:], \mathbf{X}[j,:], \mathpzc{e}_{i,j})$ & \multirow{3}{*}{$\mathbf{X}^{new}=\gamma(\mathbf{N}\mathbf{W})$} & \multirow{3}{*}{Data Sparsity, Multimodality} \\
 & $\mathbf{A}'_{i,j}=\alpha\cdot \frac{\mathbf{A}_{i,j}}{\sum_{j'}\mathbf{A}_{i,j'}}$ if $i\ne j$ else $1-\alpha$ &  &  \\
 & $\mathbf{N}=\tilde{\mathbf{D}}'^{-\frac{1}{2}}\mathbf{A}'\tilde{\mathbf{D}}'^{-\frac{1}{2}}\mathbf{X}$, where  $\tilde{\mathbf{D}}'_{i,i}=\sum_{j}\mathbf{A}'_{i,j}$ &  & \\
 \specialrule{0.5pt}{2pt}{2pt}


\multirow{2}{*}{DGRec~\cite{DBLP:conf/wsdm/Song0WCZT19}} & $\alpha_{u, v} = \frac{\mathrm{exp}(\mathbf{z}_u \cdot \mathbf{z}_v)}{\sum_{v\in N(v)}\mathrm{exp}(\mathbf{z}_u \cdot \mathbf{z}_v)}$ & \multirow{2}{*}{$\mathbf{z}_u^{new}=ReLU(\mathbf{W}\cdot\mathbf{n}_u)$} & \multirow{2}{*}{Dynamicity} \\
 & $\mathbf{n}_u = \sum_{v \in N(u)} \alpha_{u,v} \mathbf{z}_v$&  & \\
 \specialrule{0.5pt}{2pt}{2pt}

\multirow{2}{*}{KGNN-LS~\cite{wang2019knowledge}} & $\mathbf{A}'=\mathbf{A}+\mathbf{I}$ & \multirow{2}{*}{$\mathbf{X}^{new}=\gamma\big((\mathbf{N}+\mathbf{D}^{-\frac{1}{2}}\cdot \mathbf{I}\cdot \mathbf{D}^{-\frac{1}{2}}\mathbf{X})\cdot \mathbf{W}\big)$} & \multirow{2}{*}{Personalization} \\
 & $\mathbf{N}=\mathbf{D}^{-\frac{1}{2}}\mathbf{A}\mathbf{D}^{-\frac{1}{2}}\mathbf{X}$ where $\mathbf{D}_{i,i}=\sum_j \mathbf{A}'_{i,j}$&  & \\
 \specialrule{0.5pt}{2pt}{2pt}
 
\multirow{3}{*}{KGCN ~\cite{DBLP:conf/www/WangZXLG19}} & $\alpha_{u,v}^*=g(\mathbf{z}_v, \mathbf{z}_{\mathpzc{e}})$ & $\mathbf{z}_{u,sum}^{new}=\gamma\big(\mathbf{W}\cdot (\mathbf{z}_u + \mathbf{n}_u)+\mathbf{b}\big)$ & \multirow{3}{*}{---} \\
 & $\alpha_{u,v}=\frac{\exp(\alpha_{u,v}^*)}{\sum_{v'\in N(u)} \exp(\alpha_{u,v'}^*)}$ & $\mathbf{z}_{u,concat}^{new}=\gamma\big(\mathbf{W}\cdot (\mathbf{z}_u ||\mathbf{n}_u)+\mathbf{b}\big)$ &  \\
 & $\mathbf{n}_u=\sum_{v\in N(u)}\alpha_{u,v} \mathbf{z}_v$ & $\mathbf{z}_{u,context}^{new}=\gamma\big(\mathbf{W}\cdot \mathbf{n}_u+\mathbf{b}\big)$ &  \\
 \specialrule{0.5pt}{2pt}{2pt}
 
 \multirow{3}{*}{KGAT~\cite{wang2019kgat}} & $\alpha_{u,v}^*=(\mathbf{W}_{\mathpzc{e}}\mathbf{z}_v)^Ttanh(\mathbf{W}_{\mathpzc{e}}\mathbf{z}_u+\mathbf{z}_{\mathpzc{e}})$ & \multirow{3}{*}{$\mathbf{z}_u^{new}=\beta\big(\mathbf{W}_1\cdot(\mathbf{z}_u+\mathbf{n}_u)\big)+\beta\big(\mathbf{W}_2\cdot(\mathbf{z}_u\odot\mathbf{n}_u)\big)$} & \multirow{3}{*}{---} \\
 & $\alpha_{u,v}=\frac{\exp(\alpha_{u,v}^*)}{\sum_{v'\in N(u)} \exp(\alpha_{u,v'}^*)}$ &  &  \\
 & $\mathbf{n}_u=\sum_{v\in N(u)} \alpha_{u,v}\mathbf{z}_v$ &  &  \\
 \specialrule{0.5pt}{2pt}{2pt}
 
\multirow{2}{*}{DANSER~\cite{wu2019dual}} & $\alpha_{u, v} = \frac{LeakyReLU(\mathbf{w}_u(\mathbf{W}_e\mathbf{z}_{\mathpzc{e}}\odot(\mathbf{W}_p\mathbf{z_u}||\mathbf{W}_p\mathbf{z_v})))}{\sum_{v \in N(u)} LeakyReLU(\mathbf{w}_u(\mathbf{W}_e\mathbf{z}_{\mathpzc{e}}\odot(\mathbf{W}_p\mathbf{z_u}||\mathbf{W}_p\mathbf{z_v})))}$ & \multirow{2}{*}{$\mathbf{z}_u^{new}=\gamma(\mathbf{W}\cdot \mathbf{n}_u+\mathbf{b})$} & \multirow{2}{*}{Dynamicity} \\
& $\mathbf{n}_u=\sum_{v\in N(u)}\alpha_{u,v}\mathbf{z}_v$ & & \\
 \specialrule{0.5pt}{2pt}{2pt}
 
\multirow{3}{*}{RecoGCN~\cite{xu2019relation}} & $\alpha_{v,u}^r=\text{softmax}_{v}\big(\{\mathbf{W}_1^r\mathbf{z}_v\cdot \mathbf{W}_2^r\mathbf{z}_u | v\in N_r(u)\}\big)$ & \multirow{3}{*}{$\mathbf{z}_u^{new} = ReLU\big(\mathbf{W}\cdot (\mathbf{z}_u||\mathbf{n}_u) + \mathbf{b}\big)$} & \multirow{3}{*}{Scalability} \\
 & $\mathbf{n}_u^r=\sum_{\forall v\in N_r(u)} \alpha_{v,u}^r \mathbf{z}_{v}$ & & \\ & $\mathbf{n}_u=\sum_{r}\mathbf{n}_u^r $ &  & \\
\specialrule{0.5pt}{2pt}{2pt} 

PinSage~\cite{ying2018graph} & $\mathbf{n}_u=Avg/Pool\big(\{ReLU(\mathbf{W}_1\mathbf{z}_v+\mathbf{b})|v\in N(u)\}, \{\alpha_v\}\big)$ & $\mathbf{z}_u^{new}=ReLU\big(\mathbf{W}_2\cdot(\mathbf{z}_u||\mathbf{n}_u)+\mathbf{b}\big)$ & Scalability \\
\specialrule{0.5pt}{2pt}{2pt}

\multirow{2}{*}{STAR-GCN~\cite{zhang2019star}} & $\mathbf{n}_u^r = \sum\limits_{v \in N_r(u)} \frac{1}{\sqrt{|N_r(u)|\cdot|N_r(v)|}} \mathbf{W}^r \mathbf{z}_v$ & \multirow{2}{*}{$\mathbf{z}_u^{new}=\gamma\big(\mathbf{W}\cdot \gamma(\mathbf{n}_u)\big)$} & \multirow{2}{*}{Cold-Start} \\
 & $\mathbf{n}_u=\sum_{r} \mathbf{n}_u^r$ &  & \\
\specialrule{0.5pt}{2pt}{2pt} 

\multirow{2}{*}{IntentGC~\cite{zhao2019intentgc}} & $\mathbf{n}_u^r=Avg\big(\{\mathbf{z}_v| v\in N_r(u)\}\big)$ & \multirow{2}{*}{$\mathbf{z}_u^{new}=\gamma(\mathbf{W}\cdot \mathbf{z}_u+\mathbf{n}_u)$} & \multirow{2}{*}{Scalability} \\
 & $\mathbf{n}_u=\sum_{r=1}^{|\mathcal{R}|-2}\mathbf{W}^r\mathbf{n}_u^r$ &  & \\

\bottomrule[2pt]
\end{tabular}%
}
\end{table}

\subsection{Aggregator}
\subsubsection{Relation-unaware Aggregator}
For a target node $u$, a relation-unaware aggregator aims to aggregate information from parts or all of $u$'s neighboring nodes to produce a context representation. However, in this process, the relation $r=\psi(\mathpzc{e})$ ($\mathpzc{e}=e_{u,v} \text{ or } e_{v,u}$) between the target node $u$ and any neighboring node $v$ is ignored, and its information is hence not encoded in the context representation $\mathbf{n}_u$.

Fan et al.~\cite{fan2019metapath} points out that existing methods used in industry for intent recommendation rely on extensive laboring feature engineering and fail to utilize the rich interaction between users and items, which limits the model performance. To solve these issues, they model the complex objects (i.e., users and items with their attributes) as well as their interactions as a knowledge graph, and present a framework called MEIRec to learn the object embeddings for recommendation. The GNN in MEIRec generates a context representation for a target node $u$ by
\begin{equation}
    \mathbf{n}_u = g(\{\mathbf{z}_v|v\in N(u)\})
\end{equation}
where $\mathbf{z}_v$ is the embedding of node $v$ and $g$ is a aggregate function that can be either average, LSTM or CNN depending on the context. For example, they choose $g$ to be average function if $u$ is an item node, since there is usually no priority among users that have clicked/purchased an item. On the other hand, they choose LSTM as $g$ if $u$ represents a user node, because a user usually clicks items with timestamp and its neighbors can be viewed as a sequence data. 

Ying et.al.~\cite{ying2018graph} introduce a data-efficient large-scale deep recommendation engine PinSage that is developed and deployed at Pinterest. PinSage combines efficient random walks and GNN to generate embeddings of nodes that incorporate both graph structure as well as node feature information. For each target node $u$, PinSage first measures the importance of $u$'s neighboring nodes by simulating random walks starting from $u$ and computes the $L_1$-normalized visit count of nodes. Then the context representation $\mathbf{n}_u$ is computed by
\begin{equation}
    \mathbf{n}_u = Avg/Pool(\{ReLU(\mathbf{W}_1\mathbf{z}_v+\mathbf{b})|v\in N(u)\}, \{\alpha_v\})
\end{equation}
where $\mathbf{W}_1$, $\mathbf{b}$ are parameters to learn, and $\{\alpha_v\}$ are $L_1$-normalized visit counts of $u$'s corresponding neighboring nodes. 


\subsubsection{Relation-aware Subgraph Aggregator}
To handle different relations in the knowledge graph, the relation-aware subgraph aggregators split the neighborhood graph into multiple subgraphs such that all edges of a subgraph belong to only one of the relation types in $\mathcal{R}$. Each subgraph is assigned with an aggregator characterized by a unique set of parameters, and its information is extracted by this aggregator to produce a relation-sensitive context representation. We use $\mathbf{n}_u^r$ to denote the representation generated from a subgraph with type $r$ edges. Finally, $\{\mathbf{n}_u^{r_1}, \mathbf{n}_u^{r_2}, \ldots\}$ are further fused together to produce the overall context representation for target node $u$. A typical example is shown in Figure~\ref{fig:aggregator}b.


 The complex interactions in agent-initialized social e-commerce can be formulated as a knowledge graph with numerous types of relations between three types of nodes, i.e., users, selling agents and items. Xu et al.~\cite{xu2019relation} propose a novel framework RecoGCN to effectively aggregate the heterogeneous features in this knowledge graph. The ``relation-aware aggregator'' used in RecoGCN is:

\begin{equation}
    \alpha_{v,u}^r = \text{softmax}_v\big(\{\mathbf{W}_1^r\mathbf{z}_v \cdot \mathbf{W}_2^r\mathbf{z}_u|v\in N_r(u)\}\big)
\end{equation}
\begin{equation}
    \mathbf{n}_u^r = \sum_{\forall v\in N_r(u)}\alpha_{v,u}^r\mathbf{z}_v
\end{equation}
\begin{equation}
    \mathbf{n}_u = \sum_r \mathbf{n}_u^r
\end{equation}
where $\mathbf{W}_1^r$ and $\mathbf{W}_2^r$ are two learnable transformations assigned to the relation type $r \in \mathcal{R}$. 

To boost the performance in recommender systems, The STAR-GCN architecture introduced by Zhang et al.~\cite{zhang2019star} employs a multi-link graph convolutional encoder to learn the node representations. Each representation type $r\in \mathcal{R}$ is assigned with a specific transformation. Its aggregator is hence formulated as:
\begin{equation}
    \mathbf{n}_u^r = \sum_{v\in N_r(u)}\frac{1}{\sqrt{|N_r(u)|\cdot |N_r(v)|}}\mathbf{W}^r\mathbf{z}_v
\end{equation}
\begin{equation}
    \mathbf{n}_u = \sum_r \mathbf{n}_u^r
\end{equation}
where $\{\mathbf{W}^r| r\in \mathcal{R}\}$ are the parameters to learn.

IntentGC proposed by Zhao et al.~\cite{zhao2019intentgc} is a recommendation framework that captures both the explicit user preference and heterogeneous relationships of auxiliary information. To perform large-scale recommendation, IntentGC introduces a vector-wise aggregator:
\begin{equation}
    \mathbf{n}_u^r=Avg\big(\{\mathbf{z}_v|v\in N_r(u)\}\big)
\end{equation}
\begin{equation}
    \mathbf{n}_u = \sum_{r=1}^{|\mathcal{R}|-2} \mathbf{W}^r\mathbf{n}_u^r
\end{equation}
This aggregator is computationally efficient since it avoids unnecessary computations by replacing the expensive operation 
$\mathbf{W}\cdot( \mathbf{n}_u^{r_1}||\mathbf{n}_u^{r_2}||\ldots)$ 
with the summation of multiple small matrix products.


\subsubsection{Relation-aware Attentive Aggregator}
In contrast to the other two types of aggregators, relation-aware attentive aggregators convert the neighborhood graph into a weighted graph, where the weight of each edge is a function of corresponding knowledge triplet, e.g., $w=\frac{f(\mathbf{z}_u,\mathbf{z}_{\mathpzc{e}},\mathbf{z}_v)}{\sum_{v'\in N(u)}f(\mathbf{z}_u,\mathbf{z}_{\mathpzc{e}'},\mathbf{z}_{v'})}$, $w=f(\mathbf{z}_u,\mathbf{z}_{\mathpzc{e}},\mathbf{z}_v)$, etc. These weights capture the rich semantic information encoded in the edges of a knowledge graph, and measure the relatedness of different knowledge triplets to the target node $u$. 

Wang et al.~\cite{wang2019knowledge} propose an end-to-end framework KGNN-LS that utilizes a knowledge graph as an addition information source to improve recommendation. For a given user, KGNN-LS applies a trainable function to identify important knowledge graph relationships and converts the knowledge graph into a user-specific weighted graph. The aggregator in KGNN-LS is hence formulated as:
\begin{equation}
    \mathbf{A}' = \mathbf{A} + \mathbf{I}
\end{equation}
\begin{equation}
    \mathbf{N} = \mathbf{D}^{-\frac{1}{2}}\mathbf{A}\mathbf{D}^{-\frac{1}{2}}\mathbf{X}
\end{equation}
where $\mathbf{A}_{i,j}=f(\mathbf{z}_{\mathpzc{e}_{i,j}})$ and $\mathbf{D}$ is a diagonal matrix with $\mathbf{D}_{i,i} = \sum_j \mathbf{A}'_{i,j}$. $f$ is a user-specific scoring function.

Li et al.~\cite{li2019long} introduce a novel framework named V2HT, which is capable of boosting the performance of micro-video hashtag recommendation by jointly considering the sequential feature learning, the video-user-hashtag interaction, and the hashtag correlations. Similar to KGNN-LS, they assign different weights on different types of edges to transform the knowledge graph into a weighted graph. Specifically, the aggregator used in V2HT is:
\begin{equation}
    \mathbf{A}_{i,j} = f_{weight}(\mathbf{X}[i,:], \mathbf{X}[j,:], \mathpzc{e}_{i,j})
\end{equation}
\begin{equation}
\mathbf{A}'_{i,j}=
    \begin{cases}
    \alpha \cdot \frac{\mathbf{A}_{i,j}}{\sum_{j'}\mathbf{A}_{i,j'}} & \text{if $i\ne j$}\\
    1-\alpha & \text{otherwise}
    \end{cases}
\end{equation}
\begin{equation}
    \mathbf{N}=\tilde{\mathbf{D}}'^{-\frac{1}{2}}\mathbf{A}'\tilde{\mathbf{D}}'^{-\frac{1}{2}}\mathbf{X}
\end{equation}
where the subscripts $i$ and $j$ denote the $i_{th}$ and $j_{th}$ node in the graph, and $\tilde{\mathbf{D}}$ is a diagonal matrix with $\tilde{\mathbf{D}}_{i,i}=\sum_{j}\mathbf{A}'_{i,j}$. Here $\alpha$ determines the weights assigned to a node itself and other correlated nodes. When $\alpha \rightarrow 1$, the feature of a node itself will be ignored; when $\alpha \rightarrow 0$, neighboring information will be ignored. 

Fan et al.~\cite{fan2019graph} propose a novel graph neural network framework GraphRec for social recommendation. The knowledge graph in social recommendation usually consists of two graphs: a social graph denoting the relationship between users and a user-item graph denoting interactions between users and items. GraphRec introduces three aggregators to process these two different graphs, i.e., user aggregator, item aggregator and social aggregator. The user and item aggregators extract information from the user-item graph while the social aggregator distills information from the social graph. These aggregators work in a similar way. Thus we only explain the item aggregator here for simplicity. Mathematically, the item aggregator is formulated as:
\begin{equation}
    \mathbf{x}_{u,v}=MLP(\mathbf{z}_v||\mathbf{z}_{\mathpzc{e}})
\end{equation}
\begin{equation}
    \alpha_{u,v}^* = \mathbf{W}_2^T\cdot \gamma(\mathbf{W}_1\cdot (\mathbf{x}_{u,v}||\mathbf{z}_u)+\mathbf{b}_1)+\mathbf{b}_2
\end{equation}
\begin{equation}
    \alpha_{u,v}=\frac{\exp(\alpha_{u,v}^*)}{\sum_{v'\in N(u)}\exp(\alpha_{u,v}^*)}
\end{equation}
\begin{equation}
    \mathbf{n}_u = \sum_{v\in N(u)}\alpha_{u,v} \mathbf{x}_{u,v}
\end{equation}
where $\alpha_{u,v}$ represents the attention weight of the interaction with $v$ in contributing $u$'s context representation.




Recommendation in online communities faces two challenges: 1) users' interest are dynamic, and 2) users are influenced by their friends. To address these challenges, Song et al.~\cite{DBLP:conf/wsdm/Song0WCZT19} introduce a novel framework DGRec based on a dynamic graph attention neural network. Specifically, They first model dynamic user behaviors with a recurrent neural network, and its outputs are used as initial graph node embeddings. Then, a graph attention neural network is adopted to model context dependent social influence, which dynamically infers the influencers based on user's current interest. The aggregator in this graph attention neural network is:
\begin{equation}
    \alpha_{u, v} = \frac{\mathrm{exp}(\mathbf{z}_u \cdot \mathbf{z}_v)}{\sum_{v\in N(v)}\mathrm{exp}(\mathbf{z}_u \cdot \mathbf{z}_v)}
\end{equation}
\begin{equation}
    \mathbf{n}_u = \sum_{v \in N(u)} \alpha_{u,v} \mathbf{z}_v
\end{equation}

The KGCN framework proposed by Wang et al.~\cite{DBLP:conf/www/WangZXLG19} is an end-to-end recommender system that automatically discovers the high-order structure information and semantic information of Knowledge Graph (KG). The key idea of KGCN is to aggregate and incorporate neighborhood information with bias when calculating the representation for a given node in the knowledge graph. It is implemented by weighting neighbors with scores dependent on the connecting relations and nodes, which characterizes both the semantic information of KG and user's personalized interests in relations. Specifically, the aggregator in KGCN is:
\begin{equation}
    \alpha_{u,v}^* = g(\mathbf{z}_v, \mathbf{z}_e)
\end{equation}
\begin{equation}
    \alpha_{u,v} = \frac{\exp(\alpha_{u,v}^*)}{\sum_{v'\in N(u)}\exp(\alpha_{u,v'}^*)}
\end{equation}
\begin{equation}
    \mathbf{n}_u=\sum_{v\in N(u)}\alpha_{u,v}\mathbf{z}_v
\end{equation}

Wang et al.~\cite{wang2019kgat} propose a method named KGAT to explicitly model the high-order connectivities in KG in an end-to-end fashion. The aggregator in KGAT works in a similar way as that of KGCN. The only difference is that it takes all the three embeddings $\mathbf{z}_u$, $\mathbf{z}_e$, and $\mathbf{z}_v$ of a knowledge triplet as input to compute the attention score:
\begin{equation}
    \alpha_{u,v}^* = (\mathbf{W}_e\mathbf{z}_v)^T tanh(\mathbf{W}_e\mathbf{z}_u+\mathbf{z}_e)
\end{equation}
\begin{equation}
    \alpha_{u,v} = \frac{\exp(\alpha_{u,v}^*)}{\sum_{v'\in N(u)}\exp(\alpha_{u,v'}^*)}
\end{equation}
\begin{equation}
    \mathbf{n}_u=\sum_{v\in N(u)}\alpha_{u,v}\mathbf{z}_v
\end{equation}
Here $\mathbf{W}_e$ is a projection matrix to learn, which transforms inputs from the node embedding space to the edge embedding space.



In social recommendation, there are four different social effects in recommender systems, including two-fold effects in user domain, i.e., a static effect from social homophily and a dynamic effect from social influence, and the other symmetric two-fold ones in item domain. To capture these four effects, Wu et al.~\cite{wu2019dual} propose a framework named DANSER, which is composed of two dual Graph ATtention networks (GAT)~\cite{DBLP:conf/iclr/VelickovicCCRLB18}: one dual GAT for users-including a GAT to capture social homophily and a GAT to capture social influence—and the other dual GAT for items. The aggregators in DENSER work in a similar way. We show one of them as an example: 
\begin{equation}
    \alpha_{u, v} = \frac{LeakyReLU(\mathbf{w}_u(\mathbf{W}_e\mathbf{z}_{\mathpzc{e}}\odot(\mathbf{W}_p\mathbf{z_u}||\mathbf{W}_p\mathbf{z_v})))}{\sum_{v \in N(u)} LeakyReLU(\mathbf{w}_u(\mathbf{W}_e\mathbf{z}_{\mathpzc{e}}\odot(\mathbf{W}_p\mathbf{z_u}||\mathbf{W}_p\mathbf{z_v})))}
\end{equation}

\begin{equation}
    \mathbf{n}_u=\sum_{v\in N(u)}\alpha_{u,v}\mathbf{z}_v
\end{equation}
where $\mathbf{W}_p$, $\mathbf{W}_e$, and $\mathbf{w}_u$ are two weight matrices and one weight vector to learn. 


\noindent
\subsubsection{Discussion:} Compared to the other two types of aggregators, the relation-unaware aggregator loses the rich semantic information encoded in the edges of a Knowledge Graph (KG), which limits the model performance. 
 On the other hand, although the relation-aware subgraph aggregator is able to model various semantic relations in KG, it has several limitations. First, as each relation type $r\in \mathcal{R}$ is assigned with a unique set of parameters, the number of training parameters in this type of aggregators significantly increases when the relation type set $\mathcal{R}$ becomes larger. Those methods utilizing relation-aware subgraph aggregators are hence not suitable for processing knowledge graphs with numerous types of relations. Second, these aggregators process the subgraphs of different relation types independently. The correlation between different relation types are hence not encoded in the learned context representation $\mathbf{n}_u$, which limits the recommendation performance. The relation-aware attentive aggregator utilizes a single function to compute attention scores for different relation types in KG, and hence can be applied for large-scale knowledge graphs. Meanwhile, it also improves the explainability of deep recommender systems by weighting the contributions of different nodes neighboring to the target node $u$. 

\subsection{Updater}
\subsubsection{Context-only Updater} For any node $u$ in the knowledge graph, the context-only updater only receives $u$'s context representation $\mathbf{n}_u$ as input, and produces a new representation $\mathbf{z}_u^{new}=f(\mathbf{n}_u)$ for this node. Here $f$ is the updater function.

MEIRec~\cite{fan2019metapath} simply takes $\mathbf{n}_u$ as the new embedding for node $u$, i.e., $\mathbf{z}_u^{new}=\mathbf{n}_u$, which is computational efficient but may not be optimal. To overcome this issue, GraphRec~\cite{fan2019graph}, V2HT~\cite{li2019long}, DGRec~\cite{DBLP:conf/wsdm/Song0WCZT19}, KGCN ~\cite{DBLP:conf/www/WangZXLG19}, and DANSER~\cite{wu2019dual} propose to approximate $f$ via a MLP: 
\begin{equation}
    \mathbf{z}_u^{new} = \gamma(\mathbf{W}\cdot \mathbf{n}_u + \mathbf{b})
\end{equation}
where $\gamma$ is a non-linear activation function such as ReLU, and the bias $\mathbf{b}$ could be set to $\mathbf{0}$. 
STAR-GCN~\cite{zhang2019star} improves the non-linearity of $f$ by applying another activation function on $\mathbf{n}_u$ before sending it to the MLP layer:
\begin{equation}
    \mathbf{z}_u^{new} = \gamma(\mathbf{W}\cdot \gamma(\mathbf{n}_u) + \mathbf{b})
\end{equation}


\subsubsection{Single-interaction Updater}
For any node $u$ in the knowledge graph, the single-interaction updater takes both $u$'s context representation $\mathbf{n}_u$ and $u$'s current embedding $\mathbf{z}_u$ as input to compute a new representation $\mathbf{z}_u^{new}=f(\mathbf{n}_u, \mathbf{z}_u)$. $f$ is a function that involves a binary operator such as sum, concatenation, etc, which is applied to both $\mathbf{n}_u$ and $\mathbf{z}_u$. This operator builds an interaction between the node $u$ and its context, and could potentially improve the model performance.

KGNN-LS~\cite{wang2019knowledge} and KGCN ~\cite{DBLP:conf/www/WangZXLG19} both utilize summation as the interaction operator. The only difference is that KGNN-LS applies a scaling operation on the input node embeddings before feeding them to the operator. Their updaters can be written as 
\begin{equation}
    \mathbf{z}_u^{new} = \gamma\big(\mathbf{W}\cdot(\lambda\mathbf{z}_u+\mathbf{n}_u)+\mathbf{b})\big)
\end{equation}
where $\lambda$ is the scaling factor and $\mathbf{b}$ is a bias that could be set to $\mathbf{0}$. 

RecoGCN~\cite{xu2019relation} and PinSage~\cite{ying2018graph} replace the summation operator in the above updaters with the concatenation operation. It improves the expressiveness of the updater by doubling the learnable parameters, but also increases the computation cost.
\begin{equation}
    \mathbf{z}_u^{new} = ReLU\big(\mathbf{W}\cdot(\mathbf{z}_u||\mathbf{n}_u)+\mathbf{b})\big)
\end{equation}

\subsubsection{Multi-interaction Updater}
The multi-interaction updater is a combination of multiple single-interaction updaters with different binary operators, i.e., $\mathbf{z}_u^{new}=g\big(f_1(\mathbf{n}_u, \mathbf{z}_u), f_2(\mathbf{n}_u, \mathbf{z}_u), \ldots\big)$. Here $g$ is a function to fuse the representations obtained from different single-interaction updaters. This type of updaters improves feature interactions and may lead to better performance.

The only method in current literature that utilizes multi-interaction updaters is KGAT~\cite{wang2019kgat}. It considers two different operators: summation and element-wise product. The mathematical formulation of this updater is 
\begin{equation}
    \mathbf{z}_u^{new} = LeakyReLU\big(\mathbf{W}_1(\mathbf{z}_u+\mathbf{n}_u)\big)+LeakyReLU\big(\mathbf{W}_2(\mathbf{z}_u\odot \mathbf{n}_u)\big)
\end{equation}
The authors choose element-wise product because it makes the information being propagated in the graph sensitive to the affinity between $\mathbf{z}_u$ and $\mathbf{n}_u$, e.g., passing more information from similar nodes.

\subsubsection{Discussion:} Since the information of the target node $u$ is missing in the inputs of context-only updaters, they may not be sufficient to model the interaction between $u$ and its context, which limits the quality of learned representations. Single-interaction updaters address this issue by manually encoding the feature interactions between $\mathbf{z}_u$ and $\mathbf{n}_u$. Multi-interaction updaters further improve it by considering different binary operators at a time. In general, more attention is needed on the multi-interaction updaters, especially for the fusion function $g$. For example, instead of using a simple operation like summation, is it possible to learn a fusion function $g$?


\section{Solution to Practical Recommendation Issues}
\label{sec:issues}
In this section, we discuss the solution proposed by the state-of-the-art frameworks to practical recommendation issues such as scalability, data sparsity, and so on.
\subsection{Cold Start}
The cold-start problem~\cite{DBLP:journals/eswa/AcilarA09}, i.e., how to make proper recommendations for new users or new items, is a daunting dilemma in practical recommender systems. On one hand, new users and new items occupy a large portion in many real-world applications such as YouTube~\cite{DBLP:conf/recsys/CovingtonAS16}. On the other hand, the performance of recommendation largely depends on sufficient amount of historical user-item interaction data, and degrades significantly on new users/items.

\subsubsection{Uniform Term Embedding}
MEIRec~\cite{fan2019metapath} proposed by Fan et al. is an intent recommendation framework that recommends the most relevant intent (i.e., query) to a user based on his/her click history. The authors find that queries and titles of items are constituted by terms, and the number of terms are not many. So they propose to represent the queries/titles with a small number of term embeddings. It significantly reduces the number of training parameters in the model. The new queries/items that have never been searched/added before can also be represented by these terms.  

\subsubsection{Masked Embedding Training with Encoder-Decoder Architecture}
STAR-GCN~\cite{zhang2019star} adopts a multi-block graph encoder-decoder architecture. Each block contains two components: a graph encoder and a graph decoder. The graph encoder generates node representations by encoding semantic graph structure and input content features, and the decoder aims to recover the input node embeddings. To train STAR-GCN, the authors mask some percentage of the input nodes at random and then reconstruct the clean node embeddings utilizing their context information. This is referred as \textit{masked embedding training mechanism}. By using this mechanism, STAR-GCN can learn embeddings for nodes that are not observed in the training phase. In a cold start scenario, STAR-GCN initializes the embeddings of new nodes to be zero and gradually refines the estimated embeddings by multiple blocks of GNN encoder-decoders.

\subsection{Scalability}
In most existing graph neural networks, their aggregators have to visit the full neighborhood of a node to produce its node embedding, which is computationally intractable for large-scale knowledge graphs. In the following, we show some solutions proposed by the studied methods to this scalability issue.

\subsubsection{Important Node Sampling}
Instead of examining k-hop graph neighborhood to compute node embeddings, PinSage~\cite{ying2018graph} defines importance-based neighborhoods, where the neighborhood of a node $u$ is defined as the $T$ nodes that exert the most influence on node $u$. Specifically, it simulates random walks starting from $u$ and computes the $L_1$-normalized visit count for nodes visited by random walks. The neighborhood of $u$ is hence the top $T$ nodes with the highest normalized visit counts.

\subsubsection{Meta-path Defined Receptive Field}
MEIRec~\cite{fan2019metapath} and RecoGCN~\cite{xu2019relation} propose to leverage the semantic-aware meta paths to carve out concise and relevant receptive fields for each node, which is referred as meta-path defined receptive field. 

\begin{definition}\textit{\textbf{Meta-path}}~\cite{xu2019relation}.  A meta-path $\rho$ is defined as a path in a knowledge graph in the form of $t_1\xrightarrow{r_1}t_2\xrightarrow{r_2}\cdots\xrightarrow{r_l}t_{l+1}$, where there is a composite relation $R=r_1\circ r_2\circ \cdots \circ r_l$ between node type $t_1$ and $t_{l+1}$. 
\end{definition}

\begin{definition}\textit{\textbf{Meta-path defined Receptive Field (MRF)}}~\cite{xu2019relation}. Given a knowledge graph $\mathcal{G}=(\mathcal{V},\mathcal{E})$, for a node $v$ and a meta-path $\rho$ of length $l$, a meta-path defined receptive field $F_v^{\rho}=\big(f_v^{\rho}(0),f_v^{\rho}(1),\cdots, f_v^{\rho}(l)\big)$ is defined as the set of nodes that can be travelled to or passed by from node $v$ via the meta-path $\rho$, where $f_v^{\rho}(k)$ denotes the set of nodes reached by $k$ jumps on $\rho$.
\end{definition}

\begin{figure}[]
\centering
\includegraphics[width=0.5\columnwidth]{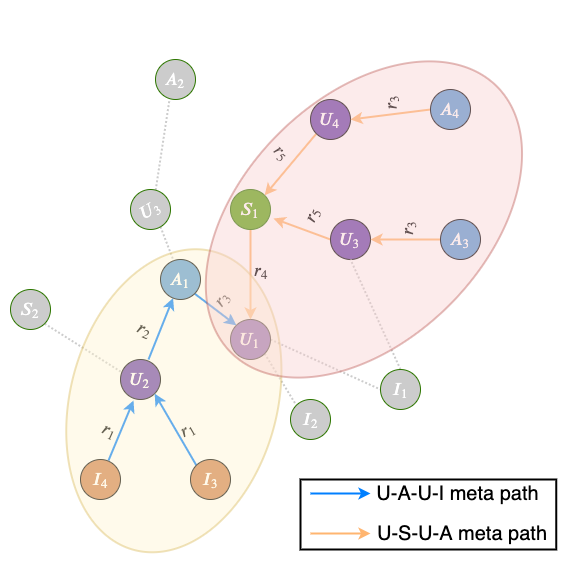}
\caption{An example of meta-path defined receptive field.}
\label{fig:mrf}
\end{figure}

Figure~\ref{fig:mrf} shows an example of MRF. Compared to $k$-hop graph neighborhood, MRF focuses only on the relations selected based on prior knowledge, and accelerates the training by greatly reducing the number of nodes in the computation graph. Moreover, it is possible to enlarge the receptive field of a node $u$ by computing multiple embeddings along different meta-paths and fusing them together to obtain a final representation: 
\begin{equation}
    \mathbf{z}_u^{new} = g(\mathbf{z}_{u, \rho_1}^{new}, \mathbf{z}_{u, \rho_2}^{new},\ldots, \mathbf{z}_{u, \rho_K}^{new})
\end{equation}
where $g$ is a fusion function. Note that the computation of these embeddings are independent from each other, and hence can be done in parallel.

\subsubsection{Vector-wise Aggregator and Updater}
To remove the limitation of training on clustered mini-graphs for large-scale graphs, IntentGC~\cite{zhao2019intentgc} introduces a special graph neural network, which replaces the normal aggregator and updater in GNNs with a vector-wise aggregator and a vector-wise updater. The key idea is to avoid unnecessary feature interactions by replacing the expensive matrix multiplication between a huge weight matrix and a giant feature vector (formed by concatenation of many vectors) with a summation of multiple small matrix products. The detailed formulation is shown in Table~\ref{tab:framework_summary}.


\subsection{Personalization}
\subsubsection{Graph Translation}
KGNN-LS~\cite{wang2019knowledge} performs personalized recommendation by converting the knowledge graph into a user-specific weighted graph, on which a graph neural network is applied to compute personalized item embeddings. The weights of edges on this new graph are computed by a trainable function $f(\mathbf{z}_e)$ that identifies important knowledge graph relationships for a given user. We refer this technique as graph translation.

\subsection{Dynamicity}
Knowledge graph could evolve over time dynamically, i.e, its nodes/edges may appear or disappear.
For example, in the scenario of social recommendation, a user's friend list may change from time to time. When a user adds a number of new friends with similar interests, the recommender system should update its recommendation strategy accordingly and reflect this change in its results.

To address this issue, Wu et al.~\cite{DBLP:conf/wsdm/Song0WCZT19} consider a dynamic feature graph setting. Specifically, for each user, they construct a graph where each node represents either this user or one of his/her friends. If user $u$ has $|N(u)|$ friends, then the total number of nodes in this graph is $|N(u) + 1|$. The node feature of friends in this graph is kept unchanged but that of user $u$ is updated whenever $u$ consumes a new item. Moreover, to capture the context-dependent social influence, the authors propose a graph attention neural network, which utilizes an attention mechanism to guide the influence propogation in its aggregator. Each friend of user $u$ is assigned with an attention weight which measures its level of influence.


\section{Datasets, Codes and Evaluations}
\subsection{Datasets}

We summarize the state-of-the-art papers and list their publicly available datasets in Table~\ref{tab:datasets} alphabetically. 
Generally speaking, we fit those datasets into 6 major scenarios: book, citation, movie, music, point of interest (POI) and social network. 
Here the general goal in each dataset is to let the recommender system infer users' preferred items based on users' past history, for example, watched movies or read books. 
The size and complexity of the datasets are upon the problem settings in different papers.

\begin{table}[t]
\caption{Summary of datasets under different scenarios}
\label{tab:datasets}
\begin{tabular}{llllll}
\toprule[2pt]
Scenario               & Dataset       & \# Entities & \# Connections & \begin{tabular}[c]{@{}l@{}}\# Relation \\ Types\end{tabular} & Citation \\
\specialrule{0.5pt}{2pt}{2pt} 
\multirow{2}{*}{Book}  &  \href{http://jmcauley.ucsd.edu/data/amazon/}{Amazon-books}  & 95,594      & 847,733        & 39                                                           &  ~\cite{wang2019camo},~\cite{zhao2019intentgc}       \\ \cline{2-6}
\addlinespace[2pt]
                       & Book-crossing & 25,787      & 60,787         & 18                                                           &  ~\cite{wang2019knowledge}, ~\cite{DBLP:conf/www/WangZXLG19}        \\
                       \specialrule{0.5pt}{2pt}{2pt} 
Citation               & S2            & -           & -              & -                                                            &  ~\cite{xiong2017explicit}        \\
\specialrule{0.5pt}{2pt}{2pt} 
\multirow{3}{*}{Movie} & Douban        & 46,423      & 331,315        & 5                                                            & ~\cite{DBLP:conf/wsdm/Song0WCZT19}, ~\cite{zhang2019star}         \\ \cline{2-6}
\addlinespace[2pt]
                       & Flixter       & 1,049,000   & 26,700,000     & -                                                            & ~\cite{zhang2019star}         \\ \cline{2-6}
\addlinespace[2pt]
                       & MovieLens    & 102,569     & 499,474        & 32                                                           &  ~\cite{palumbo2017entity2rec}, ~\cite{sun2018recurrent}, ~\cite{wang2019knowledge},  ~\cite{zhang2019star}       \\
                       \specialrule{0.5pt}{2pt}{2pt} 
Music                  & Last-FM       & 9,336       & 15,518         & 60                                                           &  ~\cite{wang2019knowledge}, ~\cite{wang2019kgat}        \\
\specialrule{0.5pt}{2pt}{2pt} 
\multirow{3}{*}{POI}   & Delicious     & 5,932       & 15,328         & -                                                            &   ~\cite{wu2019session}       \\ \cline{2-6}
\addlinespace[2pt]
                       & Yelp          & 159,426     & 6,818,026      & 6                                                            & ~\cite{DBLP:conf/wsdm/Song0WCZT19}, ~\cite{sun2018recurrent}          \\ \cline{2-6}
\addlinespace[2pt]
                       & Dianping      & 28,115      & 160,519        & 7                                                            &  ~\cite{wang2019knowledge}        \\
                       \specialrule{0.5pt}{2pt}{2pt} 
Social Network         & Epinions      & 175,000     & 508,000        & -                                                            &   ~\cite{fan2019graph}, ~\cite{wu2019dual}   \\
\bottomrule[2pt]
\end{tabular}
\end{table}

\begin{table}[]
\caption{Summary of different evaluation metrics}
\begin{tabular}{lll}
\toprule[2pt]
Evaluation Metrics & Formula & Papers \\
Metrics (@K) & & \\
\specialrule{0.5pt}{2pt}{2pt}

\multicolumn{1}{l}{Precision} & \multicolumn{1}{l}{$\frac{TP}{TP + FP}$} & \multicolumn{1}{l}{\cite{palumbo2017entity2rec}, ~\cite{sun2018recurrent}, ~\cite{wu2019dual}, ~\cite{xiong2017explicit}}\\
\specialrule{0.5pt}{2pt}{2pt}

\multicolumn{1}{l}{Recall} & \multicolumn{1}{l}{$\frac{TP}{TP + FN}$} & \multicolumn{1}{l}{\cite{li2019long}, ~\cite{DBLP:conf/wsdm/Song0WCZT19}, ~\cite{wang2019knowledge}, ~\cite{DBLP:conf/www/WangZXLG19}, ~\cite{wang2019kgat},  ~\cite{xiong2017explicit}} \\ 
\specialrule{0.5pt}{2pt}{2pt}

\multicolumn{1}{l}{F1 Score} & \multicolumn{1}{l}{$2 \cdot \frac{\textit{Precision} \cdot \textit{Recall}}{\textit{Precision} + \textit{Recall}}$} & \multicolumn{1}{l}{\cite{wang2018dkn}, ~\cite{DBLP:conf/www/WangZXLG19}} \\
\specialrule{0.5pt}{2pt}{2pt}

\multicolumn{1}{l}{MRR} & \multicolumn{1}{l}{$\frac{1}{|Q|} \sum_{i=1}^{|Q|} \frac{1}{rank_i}$} & \multicolumn{1}{l}{\cite{sun2018recurrent}, ~\cite{xu2019relation}, ~\cite{ying2018graph}, ~\cite{zhao2019intentgc}} \\
\specialrule{0.5pt}{2pt}{2pt}

\multicolumn{1}{l}{\multirow{3}{*}{NDCG}} & \multicolumn{1}{l}{$\textit{NDCG} = \frac{\textit{DCG}}{\textit{iDCG}}$} & \multicolumn{1}{l}{\multirow{3}{*}{\cite{li2019long}, ~\cite{DBLP:conf/wsdm/Song0WCZT19}, ~\cite{wang2019kgat},  ~\cite{xiong2017explicit}, ~\cite{xu2019relation}}} \\ 
\multicolumn{1}{l}{} & \multicolumn{1}{l}{$\textit{DCG} = \sum_{i=1}^{n} \frac{2^{\textit{rel}_i} - 1}{\log_2 (i+1)}$} & \multicolumn{1}{l}{} \\
\multicolumn{1}{l}{} & \multicolumn{1}{l}{$\textit{iDCG} = \sum_{i=1}^{|\textit{REL}_p|} \frac{2^{\textit{rel}_i} - 1}{\log_2 (i+1)}$} & \multicolumn{1}{l}{} \\ 
\specialrule{0.5pt}{2pt}{2pt}

\multicolumn{1}{l}{RMSE} & \multicolumn{1}{l}{$\sqrt{\frac{1}{n}\sum_{i=1}^n (\hat{y}_i - y_i)^2}$} & \multicolumn{1}{l}{\cite{fan2019graph}, ~\cite{wu2019dual}, ~\cite{zhang2019star}, } \\
\specialrule{0.5pt}{2pt}{2pt}

\multicolumn{1}{l}{HR} & \multicolumn{1}{l}{$\sum_{i=1}^{|U|} I()$} & \multicolumn{1}{l}{\cite{xu2019relation}, ~\cite{ying2018graph}} \\
\specialrule{0.5pt}{2pt}{2pt}

\multicolumn{1}{l}{MAE} & \multicolumn{1}{l}{$\frac{1}{n} \sum_{i=1}^{n} |\hat{y}_i - y_i|^2$} & \multicolumn{1}{l}{\cite{fan2019graph}, ~\cite{wu2019dual}} \\ 
\specialrule{0.5pt}{2pt}{2pt}

\multicolumn{1}{l}{\multirow{2}{*}{mAP}} & \multicolumn{1}{l}{$\textit{AP} = \frac{1}{m} \sum_{k=1}^K \textit{Precision}@k \cdot rel(k)$} & \multicolumn{1}{l}{\multirow{2}{*}{\cite{palumbo2017entity2rec}}}\\
\multicolumn{1}{l}{} & \multicolumn{1}{l}{$MAP = \frac{1}{|U|}\sum_{i=1}^{|U|}AP$} & \multicolumn{1}{l}{} \\
\specialrule{0.5pt}{2pt}{2pt}

\multicolumn{1}{l}{AUC} & \multicolumn{1}{l}{} & \cite{DBLP:conf/kdd/FanZHSHML19}, ~\cite{wang2018dkn}, ~\cite{wang2019knowledge}, ~\cite{DBLP:conf/www/WangZXLG19}, ~\cite{wu2019dual}, ~\cite{zhao2019intentgc} \\

\bottomrule[2pt]
\end{tabular}
\end{table}
\label{sec:evaluations}

\section{Future Directions}
\label{sec:future}
Although existing works have established a solid foundation for GNN-based Knowledge-Aware Deep Recommender (GNN-KADR) systems, it is still a young and promising research field. In this section, we suggest several prospective future research directions. 

\subsection{Scalability Trade-Off}
The scalability of existing GNN-KADR systems is gained at the price of sacrificing knowledge graph completeness. By sampling a fixed number of neighboring nodes, a node may lose its influential neighbors. By defining receptive field with meta-paths, the knowledge graph perceived by the model may be deprived of some important semantic information. Moreover, selecting and weighting meta-paths requires extensive prior knowledge and may be hard for some real-world applications. How to trade-off the scalability and the knowledge graph completeness is an interesting research direction. 

\subsection{Dynamicity}
Knowledge graphs are in nature dynamic in a way that nodes and edges may appear or disappear, and that node/edge inputs may change time by time~\cite{DBLP:journals/corr/abs-1901-00596}. Moreover, the types of relations between nodes in a KG may also change along time in real-world scenarios. For example, in agent-initialized social e-commerce, a user may become a selling agent to his friends someday in the future.  
Although DGRec proposed by Wu et al.~\cite{DBLP:conf/wsdm/Song0WCZT19} has partially addressed the dynamicity of graph, few of the GNN-KADR systems consider accommodating their frameworks to cases where the knowledge graph contains dynamic spatial and semantic relations. New aggregators and updaters are needed to adapt to the dynamicity of knowledge graphs. According to our literature survey, we argue that this is still an open area for future research.

\subsection{Explainability of Recommendation}
 Compared to traditional content or collaborate-filtering based recommender systems, explainability is particular important for GNN-KADR systems, because non-expert humans cannot intuitively determine the relevant context within a knowledge graph, for example, when identifying influential users in social network that are good candidates for selling agents in social e-commerce. In addition, Making explainable predictions to users allow them to understand the factors behind the network's recommendations (i.e., why was this item/services recommended?~\cite{DBLP:conf/recsys/SeoHYL17,DBLP:conf/ijcai/XiaoY0ZWC17}), and is helpful to earn user's trust on the system. It also helps the practitioner prob weights and activations to understand more about the model~\cite{DBLP:conf/www/TayTH18}.  
 
 There are several existing works focusing on the explainability of GNNs. GNNEXPLAINER~\cite{DBLP:conf/nips/YingBYZL19} proposed by Ying et al. is a model agnostic approach that provides interpretable explanations for predictions of any GNN-based model. Given an instance, it identifies a compact subgraph structure and a small subset of node features that have a crucial role in GNN’s prediction.
 Pope et al.~\cite{DBLP:conf/cvpr/PopeKRMH19} extend three common explainability methods, i.e., gradient-based saliency maps~\cite{DBLP:journals/corr/SimonyanVZ13}, Class Activation Mapping (CAM)~\cite{DBLP:conf/eccv/ZhangLBSS16}, and Excitation Backpropagation (EB)~\cite{DBLP:conf/eccv/ZhangLBSS16}, from CNNs to GNNs to identify important aspects of the computation. However, these methods are designed for homogeneous graphs and do not take the heterogeneity of knowledge graph into account.
 
 On the other hand, some other methods, e.g., KPRN~\cite{DBLP:conf/aaai/WangWX00C19}, EIUM~\cite{DBLP:conf/mm/HuangFQSLX19},  RuleRec~\cite{DBLP:conf/www/MaZCJWLMR19}, attempt to utilize the knowledge graph as an information source to make explainable recommendations. However, these methods usually sample paths from the knowledge graph and extract information from them via non-GNN algorithms such as RNN. Thus, compared to GNNs, the topological and semantic structure of knowledge graphs is corrupted in these cases, leading to unsatisfied performance.
 
 To our best knowledge, how to build an explainable GNN-based knowledge-aware deep recommender system is still an open problem and is unexplored in current literature. We believe this is the next frontier.
 
\subsection{Fairness of Recommendation}
Nowadays, recommender systems are used in a variety of domains affecting people's lives. This has raised concerns about possible biases and discriminations that such systems might exacerbate. A typical example is news recommendation, where articles with biased stand may affect people's vote decisions. 
There are two kinds of biases inherent in recommender systems~\cite{DBLP:journals/corr/abs-1809-09030}: observation bias and bias stemming from imbalanced data. Observation bias exists due to a feedback loop which causes the model only learn to predict recommendations similar to previous ones. In contrast, imbalance in data is caused when a systematic bias is present due to society, historical or other ambient biases. This bias is implicit in data so that recommender systems are usually unaware of them. 

Various methods have been proposed to build fair recommender systems. Alex et al.~\cite{DBLP:conf/kdd/BeutelCDQWWHZHC19} introduce a pairwise recommendation fairness metric to evaluate algorithmic fairness concern in recommender systems, and offer a novel regularizer to encourage improve this metric during model training. Sahin et al.~\cite{DBLP:conf/kdd/GeyikAK19} propose complementary measures to quantify bias with respect to protected attributes and present an algorithm for computing fairness-aware re-ranking of results. Specifically, their algorithm seek to achieve a desired distribution of top ranked results with respect to one or more protected attributes. Jurek et al.~\cite{DBLP:conf/www/LeonhardtAK18} propose to quantify the user unfairness or discrimination caused by the post-processing module of recommender systems, which have the original goal of improving diversity in recommendation. 

However, these algorithms are mainly designed for traditional recommender systems and cannot be used for GNN-KADR systems directly, since they do not consider any information or bias introduced by the knowledge graph. Therefore, building a fair GNN-KADR system is a promising but largely-unexplored area where more studies are expected.

\subsection{Cross-Domain Recommendation}

Besides mining a single knowledge graph, there is an increasing need of computing recommendations using data from multiple sources.
For instance, a customer may be users of more than one social networks at the same time, e.g., Facebook and LinkedIn. 
Each of these social networks collect data regarding this customer and embed them into their own knowledge graphs.
Thus, it is reasonable to leverage information from all these knowledge graphs to boost the recommendation performance.

However, there are two significant gaps in the current research.
The first gap is that existing research lacks exploration on effectively integrating information from multiple knowledge graph sources~\cite{jiang2018cross}. Most of them attempt to build associations between two different knowledge graphs by connecting related entities (i.e., nodes), as in Wang et al.~\cite{wang2017item}.
The group knowledge represented by a collection of nodes or edges of the same types, which may be critical to align multiple knowledge graphs, is ignored in these methods.

Secondly, most existing work provide cross-domain recommendations based on traditional techniques such as collaborative filtering (CF).
Some other researches use spectral clustering algorithms to aggregate knowledge from graphs of different domains, but make strong assumptions that all graphs should be available simultaneously~\cite{farseev2017cross}.
Limited effort have been made to utilize GNNs for cross-domain recommendations.
We believe this is a promising research field, since GNNs perform better than spectral clustering algorithms~\cite{kipf2016semi}.


\section{Conclusion}
In this survey, we conduct a comprehensive overview of GNN-based Knowledge Aware Deep Recommender (GNN-KADR) systems. We provide a taxonomy for their core component, i.e., the graph embedding module, which is usually a GNN in the state-of-the-art frameworks. According to the taxonomy, we categorize the aggregators of these GNNs into three groups: relation-unaware aggregator, relation-aware subgraph aggregator and relation-aware attentive aggregator. We also divide their updaters into three categories: context-only updater, single-interaction updater, and multi-interaction updater. We provide a thorough review, comparisons and summarizations of these systems within or between categories. Then, we discuss the solution proposed by these frameworks to the common practical recommendation issues such as cold-start, scalability and so on. Datasets, open-source codes and benchmarks for GNN-KADR systems are also summarized. Finally, we suggest future research directions in this rapidly growing field.
\label{sec:conclusions}

\bibliographystyle{ACM-Reference-Format}
\bibliography{bib_list.bib}

\end{document}